# Localization of narrowband single photon emitters in nanodiamonds

Kerem Bray[†, §] Russell Sandstrom[†, §], Christopher Elbadawi[†], Martin Fischer[‡], Matthias Schreck[‡], Olga Shimoni[†], Charlene Lobo[†], Milos Toth[†,*], Igor Aharonovich[†,*]

†. School of Mathematical and Physical Sciences, University of Technology Sydney, Ultimo, NSW, 2007, Australia
‡. Universität Augsburg, Lehrstuhl für Experimentalphysik IV, Universitätsstrasse 1 (Gebäude Nord), 86135 Augsburg, Germany

*Diamond nanocrystals that host room temperature narrowband single photon emitters are highly sought after for applications in nanophotonics and bio-imaging. However, current understanding of the origin of these emitters is extremely limited. In this work we demonstrate that the narrowband emitters are point defects localized at extended morphological defects in individual nanodiamonds. In particular, we show that nanocrystals with defects such as twin boundaries and secondary nucleation sites exhibit narrowband emission that is absent from pristine individual nanocrystals grown under the same conditions. Critically, we prove that the narrowband emission lines vanish when extended defects are removed deterministically using highly localized electron beam induced etching. Our results enhance the current understanding of single photon emitters in diamond, and are directly relevant to fabrication of novel quantum optics devices and sensors.*

Fluorescent nanoparticles with bright, narrowband emissions are highly sought after for applications in bio-labelling, quantum communications and sensing. Diamond nanocrystals are attracting major attention in this regard[1-9] due to their ability to host room temperature, photostable single photon emitters[10]. Several known optically active defects such as the nitrogen vacancy (NV)[11] and silicon vacancy (SiV)[12] have been studied extensively and their photophysical properties are well understood. Yet, diamond can host many other colour centres that have not been explored to date[13-14] and their origin remains unknown. While some of these centres can be engineered using ion implantation techniques[15], this avenue is not ideal, due to the damage caused by ion bombardment[16]. It is therefore highly desirable to understand emitter formation mechanisms in damage-free, scalable, bottom-up techniques such as chemical vapour deposition (CVD).

During CVD, a growing diamond crystal often incorporates silicon and other impurities[17]. This has been utilized extensively to engineer SiV[12] and other single photon emitters[18-19] in both individual nanodiamonds as well as polycrystalline films[20-21]. The process is, however, stochastic and improved control is needed over emitter concentrations and distributions. It is imperative to understand the underlying mechanisms and to achieve greater control over the incorporation of narrowband emitters in fluorescent nanodiamonds, which is essential for quantum photonic devices as well as bio-imaging applications.

Here we present three independent experiments showing that narrowband, single photon emitters in CVD-grown nanodiamonds (FWHM < 5 nm) are localized predominantly at extended morphological defects such as twin boundaries and secondary nucleation sites. Furthermore, we show that the emitters are not predominantly related to silicon impurities, contrary to suggestions made in prior literature[12]. The insights gained in this work pave the way to controlled engineering of narrowband single photon emitters in diamond.

The growth of nanodiamonds was performed using a microwave plasma CVD reactor with a hydrogen/methane ratio of 100:1 at 60 Torr, a microwave power of 900 W from 4-6 nm detonation nanodiamond seeds. To achieve reproducible identification of individual nanodiamonds by photoluminescence (PL) and scanning electron microscopy (SEM), a high resolution lithographic mask was employed after the growth. This enabled to characterize same nanocrystals in both SEM and PL.

Figure 1a,b show PL spectra recorded from individual defective and pristine nanodiamonds, respectively, grown on a silicon substrate. The insets show the corresponding SEM images of the nanodiamonds and schematic illustrations of their geometries for clarity. The nanodiamond in Figure 1a exhibits morphological defects, while Figure 1b shows a nearly-perfect nanocrystal with a cuboctahedral symmetry[22]. The PL spectrum from the nearly-perfect nanodiamond shown in Figure 1b is dominated by the SiV emission at 738 nm. A peak at 630 nm (1.967 eV) was observed in all the studied nanodiamonds. This centre is normally observed in CVD diamond films and is especially pronounced in nitrogen-doped films[13]. In the current work, we do not discuss the origin of this particular defect. The results illustrate that morphological defects correlate with the narrowband emission lines (indicated by an arrow in Figure 1a. The morphological defects include major twinning that often occurs between the <111> planes of icosahedral nanodiamonds[23] (seen as dimples between adjacent triangular facets of the nanodiamonds), and secondary nucleation sites – a common morphological defect that forms when a nanodiamond nucleates on a pre-existing crystal, forming a grain boundary along the nucleation site.

Figure 1c shows an additional spectrum recorded from a defective nanodiamond that exhibits a narrow, intense emission line at 700 nm, and Figure 1d shows a second order correlation function, $g^{(2)}(\tau)$, that confirms that the line is a single photon emitter (as evidenced by the dip below 0.5 at $g^{(2)}(0)$). The autocorrelation measurement was performed using a band-pass filter illustrated using grey dotted lines in Figure 1c. The data were corrected for background using and fit using a three level model. Most of the narrowband lines that were observed in our experiments are in fact from single photon emitters, indicating that they are point defects that are localized at the extended defects visible in SEM images.

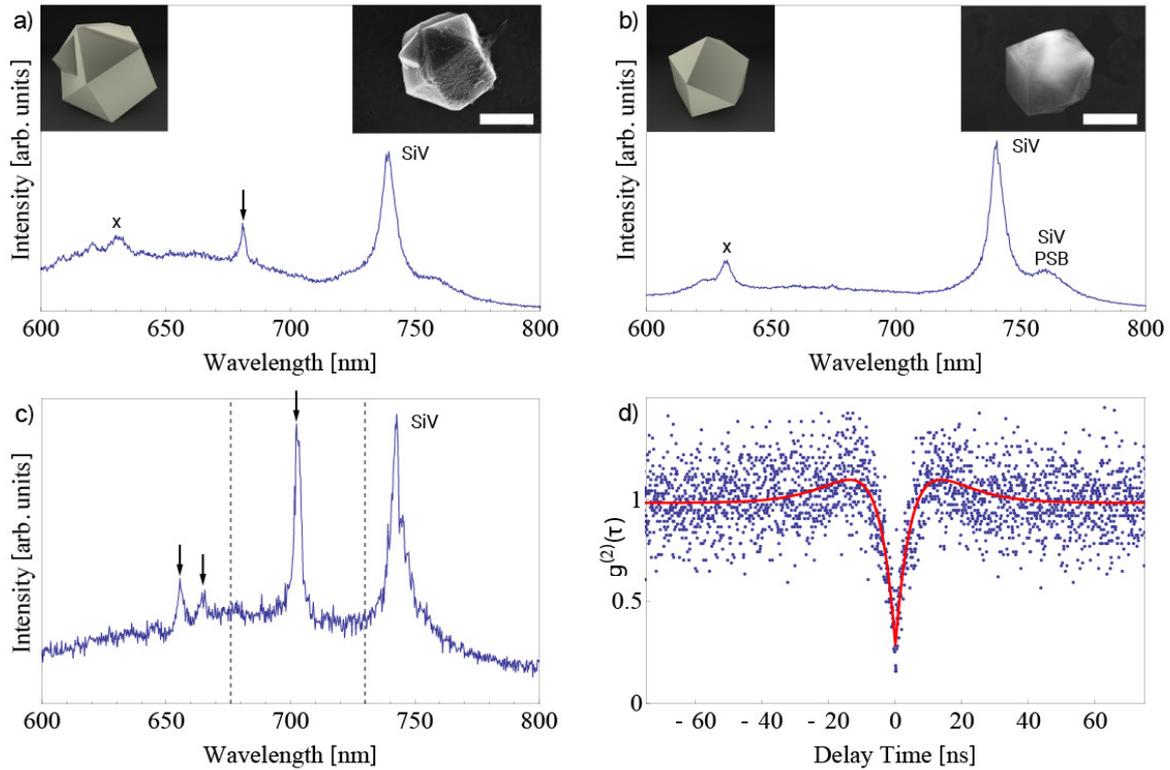

**Figure 1.** Nanodiamonds on a silicon substrate. (a) Photoluminescence spectrum from a single nanodiamond containing morphological defects. (b) Photoluminescence spectrum from a nearly-perfect nanodiamond crystal. Insets in (a,b) are SEM images of the nanodiamonds and schematic illustrations of their geometries. Scale bars in both images are 500 nm. (c) Photoluminescence spectrum from a different nanodiamond with morphological defects showing a narrow line at ~ 700 nm. (d) Second order autocorrelation function $g^2(\tau)$, recorded from the spectral region indicated by dashed lines in (c). The dip at zero delay time indicates a single photon emitter. The solid red line is a fit to the data. The samples were grown on a silicon substrate and the strong emission at 738 nm corresponds to the SiV centre. Peak assignments: SiV = Silicon vacancy, SiV PSB = Silicon vacancy phonon side band, X = CVD related peak at 630 nm, the arrows indicate narrow-band emitters studied in this work.

To ascertain whether the narrowband emitters are related to silicon impurities originating from the substrate, nanodiamonds grown on iridium were investigated. Figure shows a representative example of a nearly-perfect nanodiamond and one that contain morphological defects. A similar trend to that seen in Figure is clearly observed. The defective nanodiamond exhibit narrow sharp PL emission lines, while the nearly-perfect nanodiamond show no emission at all, or in some cases only the very weak nitrogen-related emission at 630 nm. The broadband background emission for the nanodiamond containing morphological defects is attributed to graphitic regions located at the grain boundaries. The absence of the SiV peak is expected as there was no intentional source of silicon in the growth chamber. These results indicate that the narrowband emitters are not related to silicon impurities or associated defect complexes.

A statistical survey of 60 individual nanodiamonds on both silicon and iridium substrates revealed that 90% of the investigated nanodiamonds showed a direct correlation between the presence of morphological defects and narrowband peaks. The remaining 10% of the nanodiamonds either contained morphological defects but showed no distinct narrowband emission, or they appeared to be pristine crystals and showed narrowband emissions (in these cases morphological defects may still be present but invisible in SEM images). Additional examples from both the silicon and iridium substrates are shown in the supporting information as well as the histogram of the emitters zero phonon lines and the FWHMs.

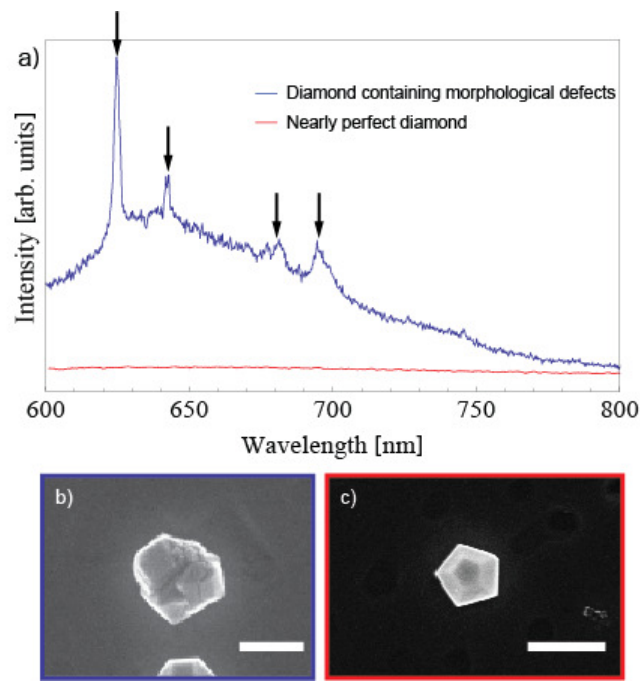

**Figure 2.** Nanodiamonds on an iridium substrate. (a) Photoluminescence spectra recorded from individual nanodiamond that do (blue) and do not (red) contain morphological defects. (b, c) Corresponding SEM images of the defective and pristine nanodiamonds. Scale bars in both images are 500 nm. The spectra were recorded using a fixed excitation power. The arrows indicate narrow-band emitters studied in this work.

The above experimental results suggest that morphological defects play a critical role in the formation of the localized narrowband emitters. We therefore performed an additional experiment that enabled us to compare PL spectra from nanodiamonds to a single crystal diamond that was grown on a diamond membrane under the same CVD conditions. The single crystal diamond membrane was lifted off type IIA single crystal diamond (Element6 Inc, [N] < 1 ppm)[24]. The membrane was placed on a silicon substrate and subjected to CVD growth under the conditions described earlier, yielding a single crystal diamond overlayer with a thickness of ~ 650 nm. Figure 3a shows a corner of the resulting overgrown membrane and a range of neighbouring nanodiamonds. The nanodiamonds were grown through spontaneous nucleation without seeds and contain multiple structural defects. The inset of Figure 3a is an optical image of the original single crystal diamond membrane.

As expected, Figure 3b shows that both the nanodiamonds and the membrane exhibit a strong SiV emission due to incorporation of silicon atoms from the substrate during growth. However, the overgrown single crystal diamond membrane does not exhibit any narrowband emission lines, while the nanodiamonds exhibit a range of narrowband emitters. This experiment confirms that the narrowband emission lines are related to morphological deformations in the nanodiamonds – likely to be twins or grain boundaries that are not present in the single crystal diamond membrane.

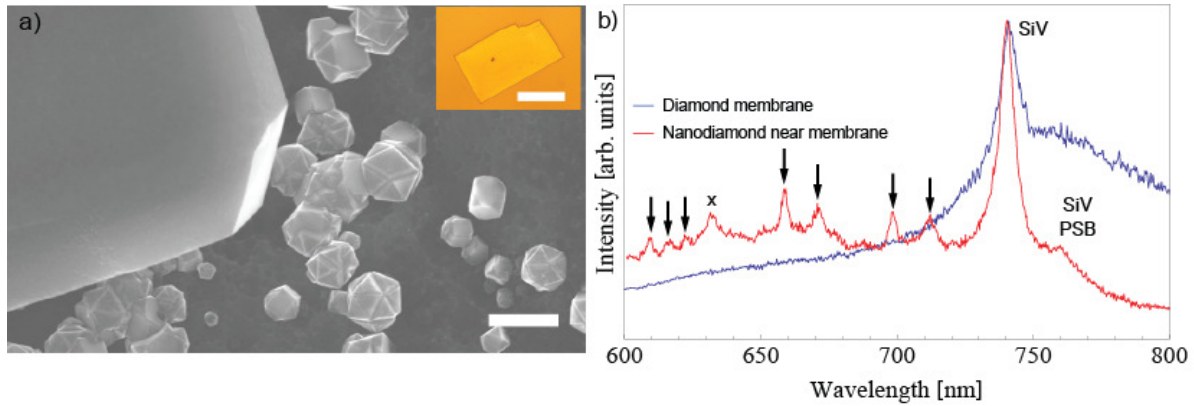

**Figure 3.** Comparison between single crystal and nanodiamond growth. (a) SEM image of a corner of the overgrown diamond membrane and nearby nanodiamonds that grew spontaneously on the silicon substrate and contain morphological defects. The scale bar is 2 µm. *Inset:* Optical image of the diamond membrane prior to growth; the scale bar is 200 µm. (b) Photoluminescence spectra recorded from the membrane and the adjacent nanodiamonds. Spectra normalised to SiV peak. Peak assignments: SiV = Silicon vacancy, SiV PSB = Silicon vacancy phonon side band, X = CVD related peak at 630 nm, the arrows indicate narrow-band emitters studied in this work.

Finally, to prove that the narrowband emitters are indeed localized at morphological defects, we select several nanodiamonds and deterministically remove the morphological defects using electron beam induced etching (EBIE) performed using water vapour as the precursor gas (Figure 4a) [25-27]. EBIE is a chemical dry etch process that is driven and localized by an electron beam. A time-lapse SEM image sequence showing EBIE of a single nanodiamond is shown in the supplementary information.

SEM images of the same nanodiamond taken before and after etching are shown in Figure 4b and figure 4c, respectively. The residual nanodiamond in Figure 4c is circled for clarity and is approximately ~ 200 nm in size. The narrowband emission lines seen in Figure 4d (blue spectrum, taken prior to EBIE) were removed by the EBIE process (red spectrum). Yet, the remaining nanocrystal still exhibits a strong SiV emission (~ 738 nm), showing that the nanodiamonds are sufficiently large to host optically active colour centres. These results solidify the hypothesis that the narrowband emissions in fluorescent nanodiamonds are point defects that decorate extended morphological defects such as twin boundaries and secondary nucleation sites.

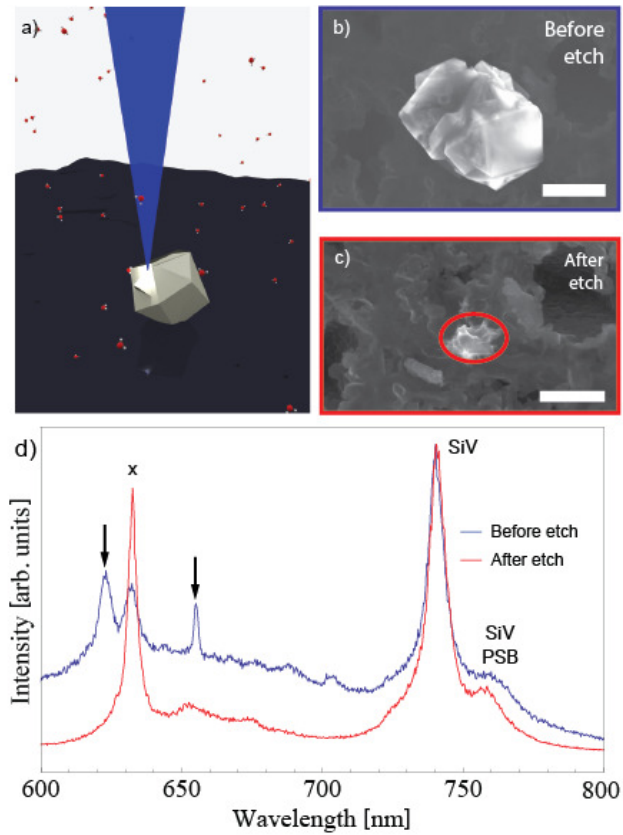

**Figure 4.** Deterministic removal of morphological defects from a single nanodiamond. (a) Schematic showing selective etching of an individual nanodiamond. (b, c) SEM images of the diamond crystal before and after electron beam induced etching. The red circle denotes the etched diamond. The contrast around the diamond is the substrate which did not fluoresce. Scale bars in both images are 500 nm. (d) Photoluminescence spectra collected before (blue curve) and after (red curve) the etch process. Spectra normalised to SiV peak. Peak assignments: SiV = Silicon vacancy, SiV PSB = Silicon vacancy phonon side band, X = CVD related peak at 630 nm, the arrows indicate narrow-band emitters studied in this work.

While we cannot deduce the chemical structure of the narrowband emitters, one can speculate about the importance of nitrogen in the formation of the defects. It is known that residual nitrogen in the gas phase modifies growth, causing twinning and re-nucleation, due changes in the growth velocities of different crystallographic planes[28]. Therefore, nitrogen atoms may also be involved in the formed isolated defects that give rise to the narrowband emission lines. Alternatively, this can result in localized deformations and stresses in the lattice[29] that may create luminescent defect states. While the origin of the emitter certainly warrants further work, the localization of narrowband quantum emitters at morphological deformations offers a fascinating opportunity for controlled localization of emitters in photonic devices as well as targeted sensing.

In conclusion, we show convincing evidence for a direct correlation between morphological defects, such as secondary nucleation sites and twin boundaries, and narrowband single photon emitters in nanodiamonds. A survey of nanodiamonds grown on silicon and iridium showed that in 90% of the cases, the presence of grain boundaries and crystal twinning correlated with the presence of narrowband emission lines in PL spectra. Single crystal diamond grown under the same conditions did not show any narrowband emitters. Finally, removal of grain boundaries from nanodiamonds using EBIE eliminated the narrowband emission lines. Our results are an important step towards understanding of origin, formation probabilities and spatial distributions of single photon emitters in nanodiamonds, enabling the deployment of nanodiamonds in bioimaging and nanophotonic technologies.

**Author information**
§ - these authors contributed equally to this work.
* (M.T.) milos.toth@uts.edu.au
* (I.A.) igor.aharonovich@uts.edu.au
**Notes**
The authors declare no competing financial interests.

**Acknowledgements**
We would like to acknowledge Mika Westerhausen for useful discussions, Stefan Lundgaard for assistance with the lithography and Andrew Magyar for assistance with membrane fabrication. The work was supported in part by the Australian Research Council (Project Number DP140102721) and FEI Company. I.A. is the recipient of an Australian Research Council Discovery Early Career Research Award (Project DE130100592). Partial funding for this research was provided by the Air Force Office of Scientific Research, United States Air Force. O.S. acknowledges the Ramaciotti foundation for financial support.